# Multimodal imaging of the mouse eye using visible light photoacoustic ophthalmoscopy and near-infrared-II optical coherence tomography

Richard Haindl[1,2,3], Valentina Bellemo[3,4], Praveenbalaji Rajendran[5], Bingyao Tan[3,4], Mengyang Liu[2,3], Qifa Zhou[6], Rainer A. Leitgeb[2], Wolfgang Drexler[2], Leopold Schmetterer[1,2,3,4,7,8,9], and Manojit Pramanik[10,*]

[1] School of Chemistry, Chemical Engineering and Biotechnology, Nanyang Technological University, 62 Nanyang Drive, Singapore, Singapore
[2] Centre for Medical Physics and Biomedical Engineering, Medical University of Vienna, Währinger Gürtel 18-20/4L, Vienna, Austria
[3] Singapore Eye Research Institute, Singapore National Eye Centre, Singapore, Singapore
[4] SERI-NTU Advanced Ocular Engineering (STANCE) Program, Nanyang Technological University, Singapore, Singapore
[5] Department of Radiation Oncology, Stanford University, Stanford, California 94305, USA
[6] University of Southern California, Department of Biomedical Engineering, Los Angeles, California 90033
[7] Ophthalmology and Visual Sciences Academic Clinical Program, Duke-NUS Medical School, National University of Singapore, Singapore
[8] Department of Clinical Pharmacology, Medical University Vienna, Vienna, Austria
[9] Institute of Molecular and Clinical Ophthalmology, Basel, Switzerland
[10] Department of Electrical and Computer Engineering, Iowa State University, Ames, Iowa 50011, USA.
* Corresponding author: Email: mano@iastate.edu

## ABSTRACT

Non-invasive imaging plays a crucial role in diagnosing and studying eye diseases. However, existing photoacoustic ophthalmoscopy (PAOM) techniques in mice have limitations due to handling restrictions, suboptimal optical properties, limited availability of light sources and permissible light fluence at the retina. This study introduces an innovative approach that utilizes Rose Bengal, a contrast agent, to enhance PAOM contrast. This enables visualization of deeper structures like the choroidal microvasculature and sclera in the mouse eye using visible light. The integration of near-infrared-II optical coherence tomography (NIR-II OCT) provides additional tissue contrast and insights into potential NIR-II PAOM capabilities. To optimize imaging, we developed a cost-effective 3D printable mouse eye phantom and a fully 3D printable tip/tilt mouse platform. This solution elevates PAOM to a user-friendly technology, which can be used to address pressing research questions concerning several ocular diseases such as myopia, glaucoma and/or age-related macular degeneration in the future.

**Keywords:** photoacoustic imaging, optical coherence tomography, ophthalmoscopy, rodent, additive manufacturing, vasculature, contrast agent, sclera

## INTRODUCTION

Non-invasive imaging has become increasingly important for enhancing diagnosis and monitoring treatment response for various organs. Emerging technologies, such as photoacoustic imaging (PAI) and optical coherence tomography (OCT), are particularly well-suited to meet the imaging needs of modern medicine as they are label-free and non-ionizing. PAI generates images by recording pressure waves produced by photon absorption of a pulsed excitation laser in chromophores [1,2], while OCT records cross-sectional tomographic images of the microstructure in biological systems and materials by measuring backscattered or back-reflected light [3].

In ophthalmology, many eye diseases involve abnormalities in the vasculature [4], making PAI a useful tool for providing detailed vascular morphological and functional parameters in addition to the scattering contrast of highly established OCT [5–7]. However, PAI and OCT face challenges when imaging deeper structures in the visible and near-infrared-I window, such as the choroid and sclera [8–

[10]. These structures are particularly important for myopia research, diagnosis, and treatment monitoring [11–14]. Current research has shown that scleral extracellular matrix remodeling is associated with axial elongation and decline in scleral strength and thickness, thus high resolution imaging of scleral structures may provide insights into the underlying mechanisms of this condition [15,16]. However, imaging of human eyes with PAI is still challenging due to their size and large water volume, dampening acoustic waves, but models for ocular diseases, like myopia, exist in mice [17].

In recent years, photoacoustic ophthalmoscopy (PAOM) has been developed for in vivo photoacoustic retinal imaging in rodents employing needle transducers [7,9,18–23], but its use in mice is limited due to handling restrictions related to the small size of the mouse eye, its suboptimal optical properties and limited permissible light fluence at the retina. This study presents an innovative and user-friendly solution to image deep structures of the mouse eye in vivo using visible light PAOM. To extend the imaging depth range beyond what is possible with existing systems, we introduced a contrast agent, Rose Bengal, which significantly enhanced the PAOM contrast and enabled visualization of the choroidal microvasculature and the sclera. To further enhance the capabilities of the system, we integrated near-infrared-II OCT to provide morphological tissue contrast and gain insight into the potential capabilities and penetration depth of future NIR-II PAOM systems.

To ensure efficient and accurate imaging, we designed an animal handling solution including a fully 3D printable tip/tilt mouse platform and a 3D printable needle transducer holder, which enable optimal placement of the transducer during imaging and effortless screening of the retina. Additionally, we developed a cost-effective 3D printable mouse eye phantom with a ball lens compatible with the transducer holder to verify system functionality, which may serve as a test phantom for small-animal PAOM systems.

This imaging platform represents a significant advancement for PAI in the field of pre-clinical ocular imaging and has the potential to be used to answer pressing research questions concerning several high-frequency and critical epidemic ocular diseases like myopia in the future.

## METHODS

### HARDWARE CONFIGURATION

The multimodal ophthalmoscope integrates absorption contrast-based PAI with scattering contrast-based OCT using a dichroic short pass mirror. A diagram of the system design can be found in Figure 1a. The pulsed PAI excitation beam is tunable within the range of 559 nm - 576 nm and has a pulse repetition rate of 1 - 5 kHz. The excitation beam, emitted by a dye laser (La1, Credo-DYE-N, Sirah dye laser, Spectra Physics, Santa Clara, CA, USA) pumped with a 532 nm diode-pumped solid-state Nd-YAG laser (INNOSLAB BX-80-2, EdgeWave GmbH, Würselen, Germany) is coupled into a single mode fiber with a numerical aperture of 0.13 (460HP, Thorlabs GmbH, Dachau, Germany). A lens-based collimator (C, 47218, EdmundOptics, Barrington, New Jersey, United States) is employed to output a beam with a beam waist diameter of 2.88 mm. A variable lens-based telescope (L3, L4, AC127-030-AB and AC127-025-AB, Thorlabs GmbH, Dachau, Germany) is utilized to compensate for refractive errors of the mouse eye, and the beam pulse energy can be adjusted up to 300 nJ with a variable density filter before fiber coupling of the excitation light of La1. Details on the fiber coupling, dye laser setup, and refractive error compensation methods can be found in previous publications [24,25].

The 1310 nm swept source OCT laser (SLE-101, Insight Photonic Solutions Inc., Lafayette, Colorado) emits fiber coupled light, which enters a fiber-based 50:50 beamsplitter (BS, TW1300R5A2, Thorlabs GmbH, Dachau, Germany) to prepare for dual balanced detection of the OCT interference pattern [26]. The two exiting fibers are connected to circulators (CC, CIR-1310-50-APC, Thorlabs GmbH, Dachau, Germany), relaying the beams to sample and reference arm respectively. The third ports of the circulators eventually connect to the final 50:50 beamsplitter, generating the interference which is detected in a dual balanced photodetector (PD, PDB470C, Thorlabs GmbH, Dachau, Germany). The collimation of the sample arm beam is performed with a triplet collimator emitting a beam with a beam waist diameter of 1.11 mm (C, TC06APC-1310, Thorlabs GmbH, Dachau, Germany). A variable lens-based telescope (L1, L2, AC254-045-C-ML and AC254-030-C-ML, Thorlabs GmbH, Dachau, Germany, Thorlabs) is used to compensate for refractive errors of the mouse eye. The reference arm comprises of the same triplet collimator and lens pairs for hardware dispersion compensation purposes [27]. A variable density filter controls the power in the reference arm, and a translational stage controls the length of the optical path.

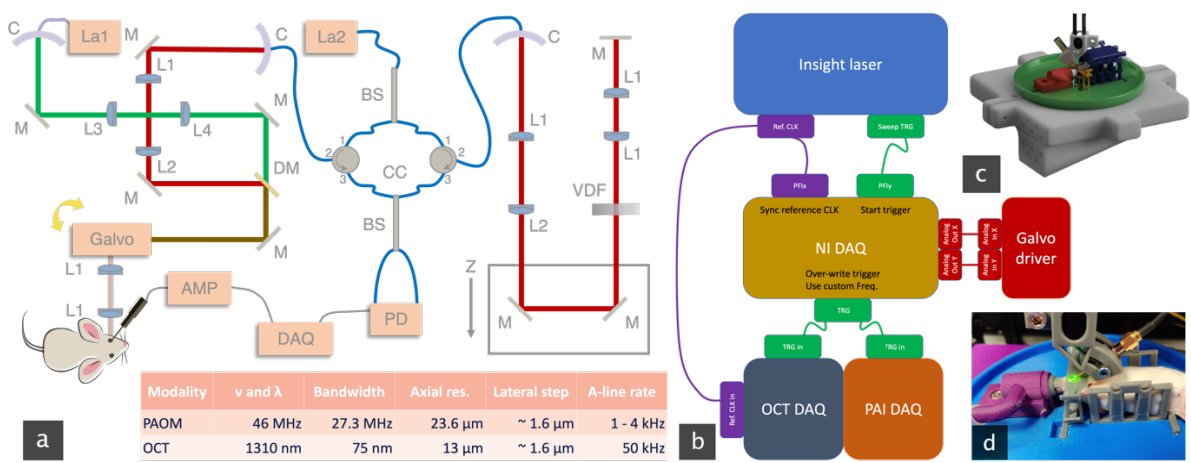

**Figure 1: Multimodal Optical Coherence Photoacoustic Ophthalmoscope (OC-PAOM)** a) Schematic of the OC-PAOM system, including the photoacoustic laser (La1) at 570 nm, OCT laser (La2) at 1310 nm, collimator (C), mirror (M), dichroic mirror (DM), lens (L), beam splitter (BS), circulator (CC), photodiode (PD), and variable density filter (VDF). The green path indicates the PAI excitation beam, the red path indicates the OCT laser beam, the brown path indicates the combined laser beam path, and the blue lines indicate the OCT fiber guide. The table lists the system specifications. b) Schematic of the OC-PAOM data acquisition and trigger control, showing the purple path for the Insight reference clock, the green path for trigger control, and the red path for galvanometric mirror control. c) 3D printable mouse holder, including the gray base plate, green half sphere, red nose cone for anesthesia, yellow headrest, blue mouse restraint unit with water heating capability, and gold needle transducer with an arc mount in gray. d) A picture of a mouse during an imaging session with the green PAOM laser and the needle transducer in contact with the eye.

A dichroic mirror (DM, DMSP1000, Thorlabs GmbH, Dachau, Germany, Thorlabs) is used to combine the PAI and OCT beam path. The combined beam is guided to the mouse eye using a telescope (L3, AC127-030-AB, Thorlabs GmbH, Dachau, Germany), to image the pivot point of the second galvanometric mirror onto the cornea of the mouse eye. The silver coated dual axis galvanometric scanners (Galvo, GVS002, Thorlabs GmbH, Dachau, Germany) raster scan the laser beam to achieve volumetric retinal imaging.

The center frequency (46 MHz) and -6 dB bandwidth (27.3 MHz) of the high-frequency needle ultrasound transducer was measured in a photoacoustic test chamber using aluminosilicate glass slides with 10 nm gold over a 2 nm Ti adhesion layer as a sample [28]. The transducer (Figure 1a, gold, connected to AMP) with an aperture size of 0.5 mm was fabricated in the Department of Biomedical Engineering of the University of Southern California using lead magnesium niobate-lead titanate (PMN-PT) as active piezoelectric material. It is connected to a wide band low power amplifier (24 dB gain, ZFL-500LN+, Mini-Circuits, New York, United States) before data readout with the PAI DAQ card.

Further information on the OC-PAOM system specifications can be found in the table of Figure 1a.

## DATA ACQUISITION AND PROCESSING

The 1310nm Insight swept OCT laser employs an internal sampling clock of 400 MHz and is configured to share a 10 MHz reference clock for synchronization purposes. This clock was used to synchronize the sampling of the 12-bit, 1 GS/s, OCT data acquisition card DAQ (ATS9371, Alazar Technologies Inc., Pointe-Claire, Quebec, Canada) with the laser sweep (Figure 1b, OCT DAQ).

The reference clock of an intermediate NI DAQ (Figure 1b, NI DAQ) card (National Instruments, PCI-6232, National Instruments Corp, Austin, Texas, United States) was synchronized to the lasers reference clock (Figure 1b, PFIx) and served as system control card. The laser's start sweep trigger was employed to synchronize the generation of a flexible A-scan trigger (Figure 1b, PFIy) with the movement of the galvanometer mirrors (Figure 1b, Analog Out X, Y).

The 16-bit PAI DAQ (M4i.4420, Spectrum Instrumentation, Ahrensfelder Weg 13, 22927 Großhansdorf, Germany) was operated at the maximum sampling rate of 250 MS/s. The internal sampling clock of the card was used for data acquisition, while the NI DAQ generated trigger (Figure 1b, TRG) was used to start the data acquisition of every A-scan and to trigger the PAI laser pulse emission.

The OCT and PAI DAQ cards were connected to the PCI-e slots and the NI DAQ to the PCI slot of a personal computer. Labview (Version 2021, National Instruments Corp, Texas, United States) was used as programming language for data acquisition and display. PAI postprocessing was implemented in Python version 3.7.14 [29] and the libraries: *matplotlib* [30], *numpy* [31], *opencv* [32] and *scipy* [33]. [28]. *En face* projections were generated with the software Fiji [34]. OCT postprocessing employed the programming language MATLAB version R2020b (MathWorks, Inc., California, United States). Standard OCT signal processing (background subtraction, resampling, and fast Fourier transform) was used to reconstruct the depth profile [35].

## TRANSDUCER ARC MOUNT

The needle transducer arc mount enables flexible angular alignment of the needle transducer with respect to the animal eye or ball lens of the mouse eye phantom, allowing high flexibility and optimization possibilities to maximize image quality and field of view for *in vivo* and phantom PAOM imaging. The transducer can be aligned between 35° and 60° towards the optical axis. The arc mount case is designed to optimize resin use with openings to the body of the case. The arc mount can be attached to a lens tube via an M6 screw and corresponding nut. The arc glider, inserted into the case, holds the transducer, which can be inserted into the top hole of the glider. An M4 screw holds the transducer in place while another M4 screw secures the arc glider within the mounting case. An exploded view of the arc mount and glider is shown in Figure 2a.

## 3D PRINTABLE MOUSE EYE MODEL

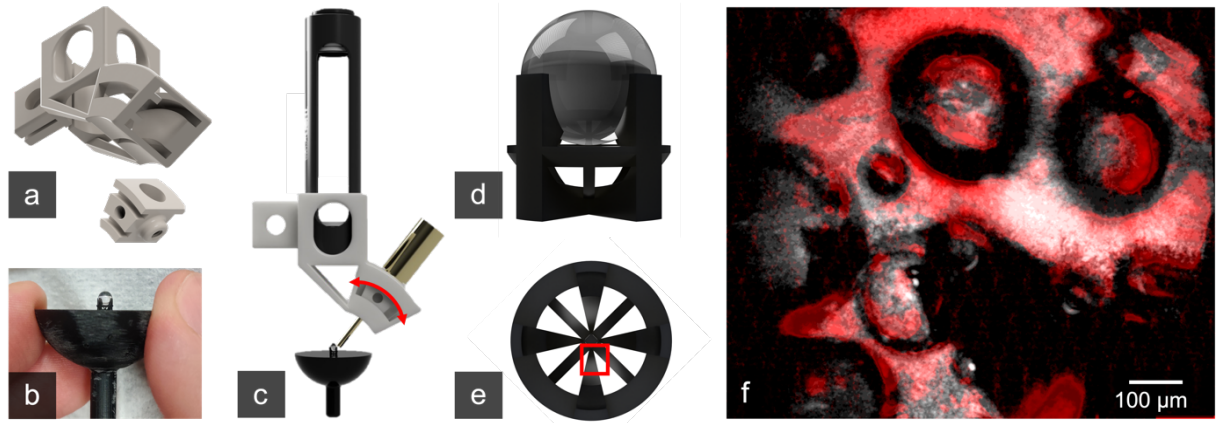

Figure 2: Transducer Arc Mount with Mouse Eye Phantom a) The transducer arc mount with a separated glider designed to hold and secure the needle transducer. b) A side view of the 3D printed and assembled mouse eye phantom. c) The fully assembled and attached arc mount with an inserted glider and needle transducer (gold), firmly fixed at the SM05 lens tube of the OC-PAOM sample arm. The needle transducer can be rotated along the focus area of the optical beam without losing contact with the ball lens (red arrow). d) A side view of the bezel with the ball lens. e) A top view of the phantom structure without the lens, where the red rectangle indicates the imaging area. f) An OCT *en face* image (gray) overlaid with a PAI *en face* image (red) with trapped air bubbles.

A fully 3D printable mouse eye model was developed for use with both imaging modalities. The model was printed in black resin (Figure 2b) with a Formlabs stereolithography printer (Form 3+, Somerville, Massachusetts, United States) and designed using Fusion 360 software (AUTODESK, San Rafael, California, United States). The base of the phantom is a half sphere with a diameter of 1 inch, and a handle is attached to the base for easy rotation. The axis of rotation is at the imaging site, allowing the needle transducer to remain in place while the phantom rotates (Figure 2c). A small N-BK7 ball lens (#43-710, EdmundOptics, Barrington, New Jersey, United States) with a diameter of 2.5 mm was secured to the phantom using a bezel mount (Figure 2d). The distance from the lens to the focal spot of 570 nm and 1310 nm beams was simulated using Zemax software. Ultrasound gel was used for coupling. The OC-PAOM imaging results demonstrated good image quality and overlap between OCT and PAI without the need for focus adjustment (Figure 2f). The image shows a multimodal *en face* projection within the red outlined region of interest in Figure 2e.

## ANIMALS

The mice utilized in the present study were subjected to considerate care and in accordance with the official ethics protocol and procedures as described therein. The Institutional Animal Care and Use Committee (NTU-IACUC) at Nanyang Technological University was the responsible local ethics body overseeing the animal experimentation. The ethical protocol for conducting animal activities was assigned the reference number A19101.

An outbred albino mouse strain of the "IcrTac:ICR, Female" variety was employed for all experimental procedures. The mice were between 8 and 16 weeks of age and were housed in an animal facility, where they were provided with environmental enrichment to reduce anxiety [36].

## IMAGING PROCEDURE

The animals were randomly selected, retrieved from the animal facility and conveyed to the imaging location directly before the start of the experiment. Subsequently, the mice were subjected to anesthesia with isoflurane (2 - 3%) and oxygen in the small rodent chamber of an anesthesia machine. Upon initial anesthesia, the mice were transferred to a homemade mouse holder equipped with a nose cone, head rest, and bite bar (Figure 1c). The anesthesia was then redirected to the nose cone to sustain the mouse's anesthetized state throughout the procedure. An artificial tear solution was utilized to moisturize the eye, and a zero-diopter contact lens (Advanced Vision Technologies, Lakewood, United States) was placed on the eyeball. The artificial tear solution was intermittently reapplied every 5 minutes to prevent cornea dehydration and cataract formation.

The holder, along with the mouse, was subsequently mounted onto the holder mount and adjusted for optimal imaging quality (Figure 1d), wherein a region of interest was preselected utilizing the fast OCT imaging system. Continuous PAI fast scans (200 x 200 pixels) were conducted to optimize the transducer position and the height of the animal holder, subsequently leading to the optimization of the imaging quality for the ensuing PAI slow scan (1000 x 1000 pixels). Parallel data acquisition is possible in this system, but data acquisition was conducted sequentially due to laser safety considerations (OCT immediately after PAI).

Following the imaging procedure, the mice were euthanized by administering an overdose of pentobarbital (100 mg/kg) via the intraperitoneal route. The euthanasia was further confirmed by cervical dislocation. The animal was eventually disposed of in accordance with the established standard procedure.

## RESULTS

### OC-PAOM IMAGING AND 3D PRINTABLE MOUSE HOLDER

The 3D printable mouse holder is composed of three primary components: a base (grey big part in Figure 1c), a half sphere (green part in Figure 1c), and attached modules (red, yellow, and blue parts in Figure 1c). The base is 270 mm x 270 mm in size, with a hole mount diameter of 170 mm at its largest point, and consists of four parts assembled with seven M6 screws. One portion of the base includes additional holes and a mounting plate for secure attachment to a translational stage (XR25P-K2/M, Thorlabs GmbH, Dachau, Germany).

The mouse holder half sphere has a diameter of 180 mm at its widest point and features engravings to accommodate the removable modules. This design allows for easy cleaning and modification of the modules, potentially saving production costs and time. The coarse parts, including the base, the half sphere, and the nose cone holder, were 3D printed with a polylactic filament (PLA) printer (Dremel 3D-20-01, Racine, Wisconsin, United States). The smaller modular parts, such as the bite bar, headrest, and mouse restraining unit, were printed using a stereolithography (SLA) printer (Form 3+, Formlabs, Somerville, Massachusetts, United States).

Figure 1d shows the assembled mouse holder during imaging, including the headrest and restraining unit. Additionally, the assembly includes a lid (not illustrated), which features a water channel for artificial heating or cooling. The headrest is secured using rubber bands, and the restraining unit is also supported by rubber bands. The lid is removable and can be attached using an M4 screws. The ICR mouse is held in the restraining unit, with its head secured in the headrest and its eye positioned at the center of rotation of the half sphere holder. The needle transducer is pointed towards the retina, and the mouse can be adjusted accordingly (rotation, tip, and tilt).

Imaging of ICR mice was performed using the 3D printed mouse mount. Figure 3 represents *en face* (Figures 3a-c) and B-scan (Figures 3d-f) imaging results of OCT and PAOM modalities. The multimodal *en face* images in Figure 3c depict the optic nerve head of the mouse and show that the two modalities are consistent in visualizing the vasculature emerging from the nerve head. PAOM provided high contrast, displaying emerging vessels in high detail, while OCT showed additional morphological contrast. With deep penetration achievable with NIR-II OCT imaging, the B-scan image in Figure 3d demonstrated OCT's ability to visualize multiple layers from the inner limiting membrane to the retinal pigment epithelium (above the red band), the choroid (red band), and the sclera below the red band [37]. The acquisition time for PAOM was longer than OCT, taking around 2

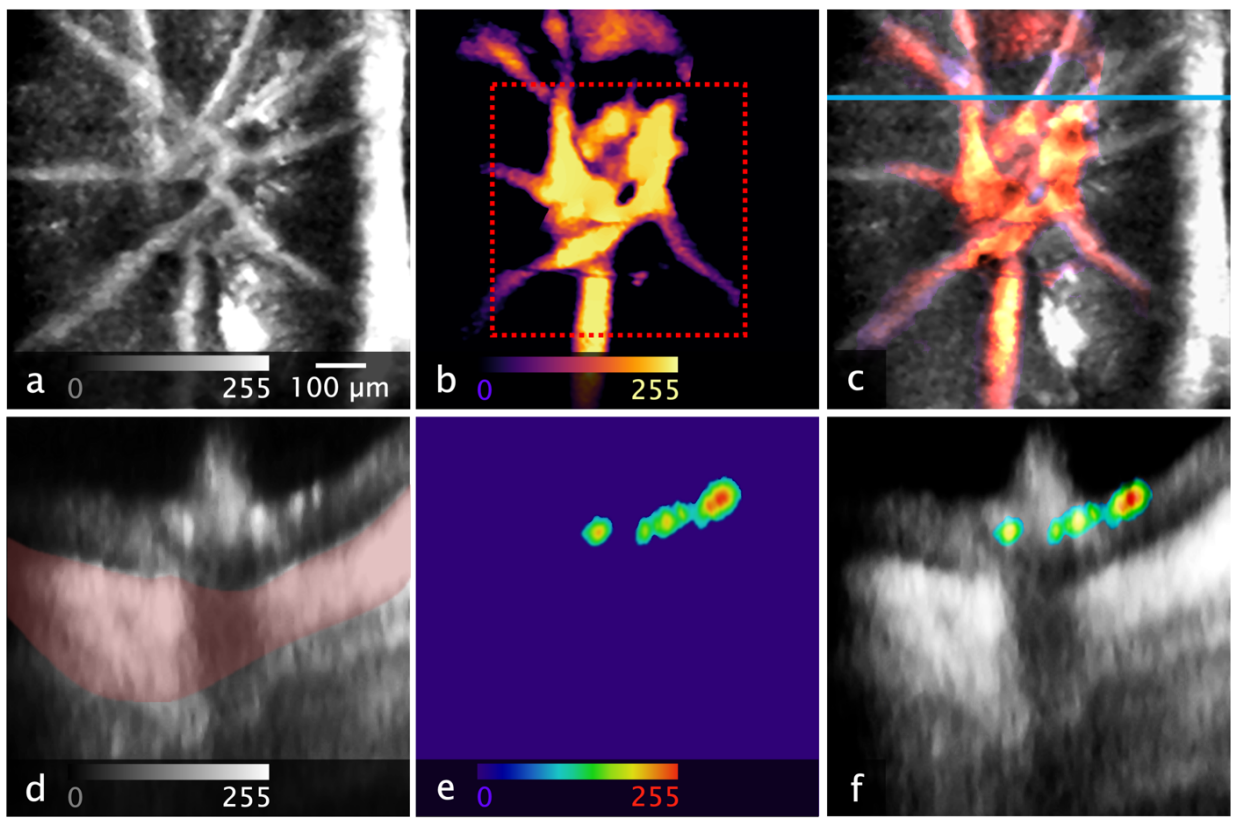

**Figure 3: Multimodal OC-PAOM Imaging Results a)** An en face average intensity OCT projection of the nerve head region. **b)** An en face maximum intensity PAOM projection of the same region. Dotted red square: Active area of the ultrasound transducer. **c)** Multimodal OC-PAOM en face projections, with PAOM in color. **d-f)** OCT, PAOM, and OC-PAOM B-scans at the indicated position in c (blue line). Red shaded area: Choroid. OCT B-Scans are median filtered and denoised.

minutes, while OCT took approximately 10 seconds.

## PAOM IMAGING WITH ROSE BENGAL

Rose Bengal is a contrast agent commonly used in ophthalmology to diagnose ocular surface disorders such as keratitis and dry eye disease. This FDA-approved diagnostic stain belongs to the xanthene class of dyes, is characterized by tetraiodo-substitution [38], and has a high absorption coefficient within the visible green light region, making it suitable for imaging procedures such as the visible light PAOM developed in this study. Although sodium fluorescein is generally preferred for angiography [39], Rose Bengal has been shown to induce collagen cross-linking [40], which is promising for myopia treatment. Therefore, we decided to evaluate the benefits of using Rose Bengal (198250, Merck KGaA, Darmstadt, Germany) for PAOM imaging quality and penetration depth (Figure 4a). We prepared different concentrations of Rose Bengal and tested them for *in vivo* imaging. A 0.1% w/v stock solution was prepared using water as a solvent, and further dilutions were made with water to reach concentrations of 0.05, 0.025, 0.0125, and 0.01% (Figure 4b). A retro-orbital sinus injection was then performed to apply a specified volume of the dye solution to the retro-orbital sinus, which circulated in the retinal vasculature of the mouse [41]. We applied a maximum volume of 150 µl.

Injecting 150 µl of 0.1% Rose Bengal produced the most promising outcomes as opposed to lower concentrations, which did not improve the contrast as compared to hemoglobin. B-scan images taken after a retrobulbar injection of 0.1% Rose Bengal showed increased acoustic intensity and absorption not only in the retina but also in the choroid (Figure 4c) and sclera (Figure 4d, green marks), allowing for the visualization of small deep structures.

Furthermore, retinal imaging with Rose Bengal showed an expanded field of view beyond the transducer's active area (500 µm x 500 µm) and allowed for the visualization of small vessels connected to larger branches in the maximum intensity *en face* projection of the retinal layer (Figure 4e). The boundary between retinal and choroidal vasculature was also detectable, indicating high axial resolution with the high-frequency sensor technology (Figure 4c). Additionally, a maximum intensity *en face* projection of the choroidal layer allowed for the visualization of the capillary system in the choroid with a resolution down to a few-micrometers, along with larger vessels (Figure 4d and f).

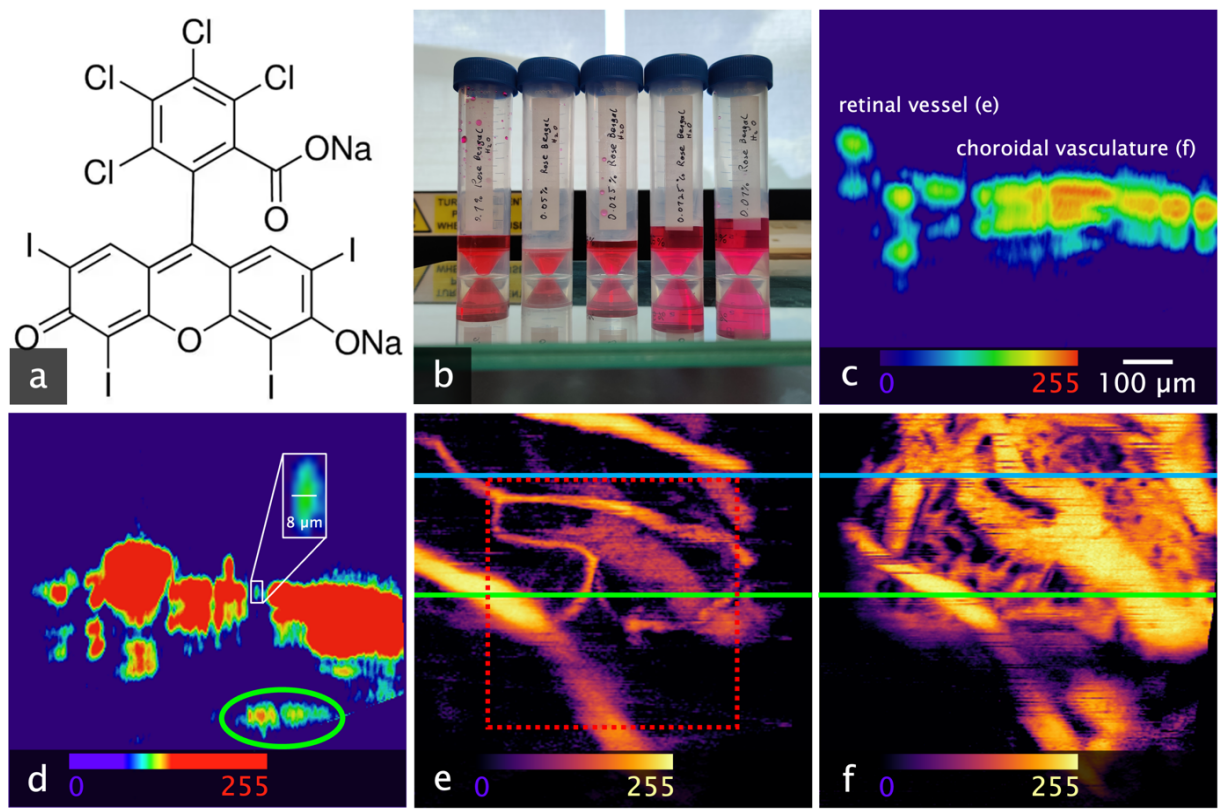

**Figure 4: Rose Bengal PAOM Imaging a)** The chemical structure of Rose Bengal disodium salt. **b)** Dye preparation from powder with various concentrations of Rose Bengal (0.1%, 0.05%, 0.025%, 0.0125%, and 0.01% from left to right). **c)** A typical PAOM B-scan (blue line in e and f). **d)** Contrast adjusted PAOM B-scan (green line in e and f) with unobstructed scleral features (green ellipse). **e, f)** *En face* projections at different depth for the retinal and choroidal layers, respectively, with the scale bar valid for all images. Dotted red square: active area of the ultrasound transducer.

Notably, the choroidal microvasculature was only visible with Rose Bengal injection, highlighting its potential advantage for PAOM. Based on these findings, using the optimized parameters for Rose Bengal injection provides an effective tool for non-invasive retinal and choroidal imaging with high resolution and contrast.

## DISCUSSION

Ocular rodent animal imaging is crucial in pre-clinical research, as the model is highly similar to humans [42–44]. The use of imaging techniques provides a unique opportunity to study and enhance our knowledge of various diseases. Here, we presented a novel concept for a rodent holder that enabled rotation without repositioning of the rodent's translational position. This allows imaging with the limited field of view accessible with PAOM. Additionally, we designed a needle transducer holder that met the stringent positional and alignment requirements of the transducer and excitation beam path. We demonstrated the fully 3D printable holder with imaging results obtained by multimodal OC-PAOM imaging of the mouse eye's retina, choroid, and sclera with and without exogenous contrast. With PAOM, both the nerve head region and highly detailed choroidal vasculature can be imaged, while the retina, choroid, and sclera were accessible with NIR-II OCT, indicating the feasibility for NIR-II PAOM to penetrate through the choroid with the lower vascular absorption and to study scleral disease with the aid of contrast agents such as gold nanorods [45] in the future. Further, Rose Bengal proved to be a potent dye for PAOM for green excitation light, allowing for a choroidal vascular and sclera visualization in mice with PAOM for the first time.

We have chosen Rose Bengal as contrast agent due to its noteworthy spectroscopic and photochemical properties. Rose Bengal has a strong propensity for intersystem crossing, which leads to the generation of a photochemically active triplet excited state [38], when exposed to laser light. This process leads to the generation of reactive oxygen species (ROS). The ROS can react with collagen fibril molecules, promoting the formation of new chemical bonds between amino groups through photopolymerization, which can increase the mechanical strength of the tissue and enhance its stiffness. This process has shown promise in increasing the stiffness of scleral tissue, which may have implications for the treatment of myopia [46]. Therefore, Rose Bengal is a versatile, theranostic contrast agent and the demonstration of its effectiveness in PAOM opens new avenues to study ocular diseases.

In literature, PAOM has been used to study mice, rats, and rabbits as ocular models [47,48]. Although mice and rats have small eyes, rabbits are closer in size to human eyes. However, rabbits lack a fovea (like rodents), and their retinal vasculature differs from that of humans and rodents. Moreover, rabbits have a unique horizontal streak of myelinated retinal nerve fiber layer called the medullary ray that is absent in humans and rodents [49]. Nonetheless, rabbits have been successfully used to study ocular diseases such as choroidal vascular occlusions (CVO), utilizing a multimodal PAOM and OCT imaging system which allows for visualization, characterization, and quantification of laser-induced CVO with high resolution and contrast [50]. Similarly, integrated PAOM and OCT systems have been used to study retinal and choroidal vessels, retinal vein occlusion, and retinal neovascularization in rabbits [50–55]. In rats, PAOM has been used in combination with scanning laser ophthalmoscopy, optical coherence tomography, and fluorescein angiography to visualize the choroidal vasculature in albino rats, but not in pigmented rats [21,56,57]. Additionally, PAOM with NIR-I excitation has been used to visualize the choroidal vasculature, but the quality of the imaged retinal vasculature in rats was lower compared to visible light excitation [9]. Multi-wavelength illumination was also used to quantify the metabolic rate of oxygen [19]. However, there is limited research on mice using PAOM [20], and thus far, PAOM has not been able to resolve the choroidal vasculature nor the sclera in mice.

NIR-II OCT is rarely used for retinal imaging due to its high water absorption and low axial resolution. Literature reports are scarce due to those challenges [58–61]. However, our system uses a high-power swept source, allowing us to overcome the limitations of NIR-II OCT and we were therefore able to image deep and with high contrast. This enables direct observation of the sclera without surgery or imaging from the side of the eye globe, offering an advantage for both throughput, long-term studies, and animal welfare.

In the future, we plan to replace the PAOM excitation laser with a faster model that offers a higher tunability of the wavelength range and access to longer wavelengths for deeper imaging of the sclera. We will investigate different exogenous contrast agents to enhance scleral contrast in myopic mouse models to further our understanding of myopia disease onset and progression. Furthermore, we will target scleral cross-linking, recently demonstrated with green light [62], combining treatment and monitoring with PAOM using the same light source.

## CONCLUSION

Our study provides a comprehensive solution for imaging deeper structures of the mouse eye *in vivo*, by combining PAOM with contrast agents, integrating NIR-II OCT, and addressing animal handling restrictions with a novel 3D printable platform and phantom. The ocular imaging platform has the potential to be used to answer pressing research questions concerning several high-frequency and critical epidemic ocular diseases like myopia in the future.

## FUNDING

This work was funded by grants from the National Medical Research Council (CG/C010A/2017_SERI; OFLCG/004c/2018-00; MOH-000249-00; MOH-000647-00; MOH-001001-00; MOH-001015-00; MOH-000500-00; MOH-000707-00; MOH-001072-06; MOH-001286-00), National Research Foundation Singapore (NRF2019-THE002-0006 and NRF-CRP24-2020-0001), A*STAR (A20H4b0141), the Singapore Eye Research Institute & Nanyang Technological University (SERI-NTU Advanced Ocular Engineering (STANCE) Program), and the SERI-Lee Foundation (LF1019-1) Singapore.

M. Liu is funded by H2020-MSCA-IF-2019 project SkinOptima with grant agreement ID 894325.

## ACKNOWLEDGEMENT

We thank Bei Shi Lee for input and advice as well as proofreading of the manuscript. We also like to thank Rhonnie Austira Dienzo for useful discussion and help regarding animal imaging.